\documentclass[useAMS,usenatbib]{mn2e}
\usepackage{epsfig}
\usepackage{amsmath}
\usepackage{rotating}

\newcommand{\be}{\begin{equation}}
\newcommand{\beq}{\begin{equation}}
\newcommand{\ba}{\begin{eqnarray}}
\newcommand{\ee}{\end{equation}}
\newcommand{\eeq}{\end{equation}}
\newcommand{\ea}{\end{eqnarray}}

\newcommand{\hs}{\hspace{1mm}}

\newcommand{\apj}{ApJ}
\newcommand{\aap}{A\&A}
\newcommand{\apjl}{ApJL}
\newcommand{\mnras}{MNRAS}
\newcommand{\aj}{AJ}
\newcommand{\apjs}{ApJS}

\def\lsim{~\rlap{$<$}{\lower 1.0ex\hbox{$\sim$}}}

\def\gsim{~\rlap{$>$}{\lower 1.0ex\hbox{$\sim$}}}

\title[Upper Limits on Large Dust Grains from the SXB]{Upper Limit on
Dimming of Cosmological Sources by Intergalactic Grey Dust from the
Soft X-ray Background}

\author[Mark Dijkstra \& Abraham Loeb]{Mark
Dijkstra\thanks{E-mail:mdijkstr@cfa.harvard.edu} and Abraham
Loeb\thanks{E-mail:aloeb@cfa.harvard.edu}\\ Institute for Theory \&
Computation, Harvard University, 60 Garden Street, Cambridge, MA
02138, USA}

\begin{document}

\date{\today} \pagerange{\pageref{firstpage}--\pageref{lastpage}}
\pubyear{2006}

\maketitle
\label{firstpage}
\begin{abstract}
Active Galactic Nuclei (AGN) produce a dominant fraction ($\sim 80\%$)
of the Soft X-ray background (SXB) at photon energies $0.5<E<2$ keV.
If dust pervaded throughout the intergalactic medium, its scattering
opacity would have produced diffuse X-ray halos around AGN. Taking
account of known galaxies and galaxy clusters, only a fraction $F_{\rm
halo}\lsim 10\%$ of the SXB can be in the form of diffuse X-ray halos
around AGN. We therefore limit the intergalactic opacity to
optical/infrared photons from large dust grains (with radii in the
range $a=0.2-2.0\mu$m) to a level $\tau_{\rm GD}\lsim 0.15(F_{\rm
halo}/10\%)$ to a redshift $z\sim 1$.  Our results are only weakly
dependent on the grain size distribution or the redshift evolution of
the intergalactic dust.  Stacking X-ray images of AGN can be used to
improve our constraints and diminish the importance of dust as a
source of systematic uncertainty for future supernova surveys which
aim to improve the precision on measuring the redshift evolution of
the dark energy equation-of-state.
\end{abstract}

 
\section{Introduction}
\label{sec:intro}
The use of Type Ia supernovae as standardized candles provides a
powerful measure of the redshift dependence of the luminosity
distance, and therefore of the cosmological parameters that shape our
Universe \citep[e.g.][and references therein]{W07,R07}.  Historically,
supernova surveys provided the first robust evidence for the existence
of a positive cosmological constant \citep{R98,S98}. This inference
has been confirmed by other independent probes, such as the location
of the acoustic peaks in the Cosmic Microwave Background
\citep{wmap5,wmap} and galaxy surveys \citep{Eisenstein,E2}, and the
observed evolution of the mass function of galaxy clusters \citep{V08}.

Since supernovae measure the luminosity distance at
optical/infrared-infrared wavelengths, they suffer from a systematic
uncertainty owing to the possible existence of intergalactic dust. In
particular, 'grey' dust which consists of large ($>0.1\mu$m) grains
that may preferentially reside in the intergalactic medium (IGM), can
suppress the observed flux while producing little reddening
\citep{A99A,A99B,BF05}. Therefore, the presence of intergalactic grey
dust could lead to a systematic overestimation of the luminosity
distance, which would in turn modify the inferred values of the
cosmological parameters. This becomes an important source of
systematic uncertainty for future ambitious supernova surveys
(performed e.g. on the  Large Synoptic Survey Telescope
(LSST)\footnote{http://www.lsst.org}, or by the The Joint Dark Energy
Mission (JDEM)\footnote{http://jdem.gsfc.nasa.gov/}) which are
designed to constrain the equation of state parameter of the dark
energy, $w(z)$, to an unprecedented (percent level) precision. A
systematic suppression of the supernova flux by a small amount of
$+\Delta m$ magnitudes due to grey dust would result in a best fit $w$
that is systematically offset by $\Delta w\sim -2\Delta m$
\citep{Zhang08}.

Probing the grey dust content of the IGM is not only important for the
purpose of limiting a source of systematic error for future supernova
surveys. It is also related to the fundamental problem of how heavy
elements were first produced in galaxies and then and then dispersed
through outflows into the intergalactic medium
\citep[e.g.][]{H00,A01,FL03,BF05,S06,Dave07,O8}. A measurement of the
intergalactic dust abundance would serve as an important new
constraint on theoretical models of the enrichment process of the IGM
\citep[e.g.][]{L97,AH00,I04,P06}.

In this paper we constrain the abundance of intergalactic grey dust
using two facts about the Soft X-ray Background (SXB): {\it (i)}
Active Galactic Nuclei (AGN) account for $\sim 80\%$ of the total SXB
\citep[][also see \S~\ref{sec:result}]{H06,H07b}; and {\it (ii)} the
unresolved SXB component is partially accounted for by the known
population of galaxies and X-ray clusters \citep[][and see
\S~\ref{sec:sxb}]{M03,H07}.  If large dust grain existed in the IGM,
they would have scattered X-rays by sufficiently large angles to
create extended X-ray halos around point sources \citep[e.g.][and
references therein]{AH78,M99,P06}. The lack of such a halo around any
particular point source can be used to limit the grey dust abundance
along its line-of-sight. Indeed, the absence of an X-ray halo around a
$z=4.3$ quasar allowed \citet{P06} to place an upper limit on the
intergalactic abundance of dust grains with a radius of $a=1\mu$m. In
this paper, we apply this pioneering approach to the SXB as a
whole. The advantage of our methodology is that it does not require
knowledge of the surface brightness profile for individual X-ray
halos, which quite  strongly depends on the assumed redshift evolution
and size distribution of the intergalactic dust grains (see
\S~\ref{sec:discuss}).

In \S~\ref{sec:method} we describe the model used to place upper
limits on the intergalactic opacity in grey dust. In
\S~\ref{sec:result} we present our numerical results, whose
implications are further discussed and compared to previous work in
\S~\ref{sec:discuss}.  Finally, we summarize our final conclusions in
\S~\ref{sec:conc}. Throughout our discussion, we adopt the standard
set of cosmological parameters for the background cosmology,
$(\Omega_m,\Omega_{\Lambda},\Omega_b,h,\sigma_8)=(0.27,0.73,0.042,0.70,0.82)$
\citep{wmap5}.

\section{Constraints from the SXB}
\label{sec:method}

\subsection{The Composition of the SXB}
\label{sec:sxb}
The total SXB in the 1--2 keV band amounts to $4.6\pm 0.1\times
10^{-12}$ erg cm$^{-2}$ s$^{-1}$ deg$^{-2}$ \citep[e.g.][their
Fig.~15]{H06}. As already mentioned, observed AGN account for $\sim
80\%$ of this flux \citep[][also see
\S~\ref{sec:result}]{H06,H07b}. \citet{H07} showed that starburst and
'normal' galaxies that are too faint in X-rays to be resolved, account
for $\sim 10$--$15\%$ of the SXB. Furthermore, as much as $\sim 6-9\%$
may be accounted for by spatially extended X-ray sources such as
clusters \citep[e.g.][but see Hickox \& Markevich
2006]{Wu,M03,D04}. Lastly, 1.0-1.7\% of the SXB must consist of
Thomson scattered X-rays that were originally emitted by AGN
\citep{Soltan03}. Adding up these known contributions could, in
principle, account for the full SXB. As a conservative working
hypothesis, we therefore assume that X-ray halos -- produced by dust
scattering around AGN, cannot account for more than a fraction $F_{\rm
halo}\lsim 10\%$ of the total SXB.

We do not simply adopt the unresolved X-ray background that has been
derived by others: \footnote{\citet{H07} estimate the flux in the
unresolved SXB to be $3.4\pm 1.4\times 10^{-13}$ erg cm$^{-2}$
s$^{-1}$ deg$^{-2}$, which is $\sim 7\pm 3\%$ of the total. Similarly
\citet{M03} claim that $6 \pm 6 \%$ of the SXB in the energy range
$E=0.5-2.0$ keV is unresolved.} \citep[e.g.][]{H07}.
When \citet{H07} measure the flux in the 'unresolved' component they
exclude regions around detected X-ray sources of radius $\lsim 4-9
r_{90}$, in which $r_{90}=2.2$ arcsec denotes the radius of the region
in which $90\%$ of the flux of a point source is detected. X-ray halos
that are a result of scattering off intergalactic dust are expected to
be $\sim 0.5-2$ arcminutes in diameter (see \S~\ref{sec:Xscat}), which
can be comparable to the size of the excluded regions. Therefore, a
significant fraction of X-rays that were scattered by intergalactic
dust would already be excluded from the measurement of the unresolved
SXB. On the other hand, when measuring the soft X-ray luminosity from
AGN, the X-ray flux from within $r_{90}$ is used, which contains a
negligible fraction of X-rays that were scattered by intergalactic
dust.

\subsection{The Scattering Cross-Section of Dust Grains}
\label{sec:Xscat}

The cross-section for scattering of radiation by a dust grain can be
written in the form, $\sigma_{\rm scat}=Q_{\rm scat} \pi a^2$, where
$a$ is the radius of the grain (assumed to be spherical). Here,
$Q_{\rm scat}$ denotes the scattering efficiency, which is defined as
the ratio between the geometric cross-section and the absorption
cross-section of the grain, and is given by \citep{AH78,M99},
\begin{equation}
Q_{\rm scat} \approx \left\{ \begin{array}{ll} \ 0.7(a / \mu{\rm m})^2
         (6 \hs {\rm keV}/E)^2 & \mbox{$Q_{\rm scat} < 1$};\\ \ 1.5 &
         \mbox{otherwise}.\end{array} \right.
\label{eq:Q}
\end{equation} 
The discontinuity at $Q_{\rm scat}=1$ occurs when the electromagnetic
phase shift across the grain reaches unity \citep{AH78,M99}. For a
fixed grain size, Eq.~(\ref{eq:Q}) yields a constant cross-section up
to a fixed threshold in photon energy $E$, after which it declines as
$E^{-2}$. All dust grains with radii $a\gsim 0.16(E/1\hs{{\rm
keV}})^{2}\mu$m are equally likely to scatter all photons of energy
less than $E$. This important result allows us to place tight
constraints on intergalactic dust. Equation~(\ref{eq:Q}) does not
apply to arbitrarily low photon energies, and $Q_{\rm scat}$ typically
decreases drastically below unity at $E \lsim 1(a/0.1\mu{\rm m})^{-1}$
eV for grains of radius $a$ \citep[see e.g. Fig~2-4 of][]{LD93}.

X-ray scattering by large dust grains can be described by a phase
function of the form $P(\theta_{\rm scat})\propto \exp(-\theta_{\rm
scat}^2/2\sigma^2)$, in which $\sigma=1.04(a/\mu m)^{-1}(E/{\rm
keV})^{-1}$ arcmin \citep{MG86}. Furthermore, $\theta_{\rm scat}$
denotes the angle between the propagation direction of the photon
before and after scattering (i.e. ${\bf k}_{\rm in}\cdot{\bf k}_{\rm
out}\equiv \cos[\theta_{\rm scat}]$). Hence, dust typically scatters
X-ray photons forward into a halo around the source of angular size
$\sim 2\sigma$.

\subsection{The Fraction of X-rays that is Scattered by Dust}

The fraction of X-ray photons {\it observed} at energy $E$ that is
expected to be scattered into X-ray halos by intergalactic dust is
given by,
\begin{equation}
F(E)=\int_{0}^{\infty}dz'\mathcal{F}(z')[1-e^{-\tau_{\rm
GD}(E,z')}]\Big{/}\int_{0}^{\infty}dz'\mathcal{F}(z') ,
\label{eq:fe}
\end{equation}
where we have defined the function $\mathcal{F}(z)\equiv
\mathcal{L}(z)/(1+z)^2\mathcal{E}(z)$, in which
$\mathcal{E}(z)=\sqrt{\Omega_m(1+z)^3+\Omega_{\Lambda}}$, and
$\mathcal{L}(z)$ denotes the X-ray volume emissivity (in erg s$^{-1}$
cMpc$^{-3}$, see \S~\ref{sec:xem}).

The total optical depth to X-ray scattering by grey dust between
redshift $0$ and $z$ for an observed photon energy $E$ is given by
\begin{equation}
\tau_{\rm GD}(E,z)=\frac{c}{H_0}\int_{0}^{z}dz'\frac{n(z')\sigma_{\rm
scat}(E,z')(1+z')^2}{\mathcal{E}(z')},
\label{eq:tau1}
\end{equation} 
where $n(z)$ is the comoving number density of dust grains at a
redshift $z$, each having a cross section $\sigma_{\rm
scat}(E,z)=Q_s(E\times[1+z])\pi a^2$ with $Q_s$ given by
Eq.~(\ref{eq:Q}). Equation~(\ref{eq:tau1}) implicitly assumes that all
dust grains are of the same size. For a distribution of grain sizes,
Eq.~(\ref{eq:tau1}) is generalized to the form,
\begin{eqnarray}
\tau_{\rm GD}(E,z)=\frac{c}{H_0}\int_{0}^{z}dz' \int_{a_{\rm
min}}^{a_{\rm max}}da\frac{dn}{da}\frac{\sigma_{\rm
scat}(E,z',a)(1+z')^2}{\mathcal{E}(z')},
\label{eq:tau2}
\end{eqnarray}
where $\frac{dn}{da}da$ denotes the comoving number density of dust
grains with radii in the range $a\pm da/2$.

\subsection{The X-ray Volume Emissivity}
\label{sec:xem}

\begin{figure}
\vbox{\centerline{\epsfig{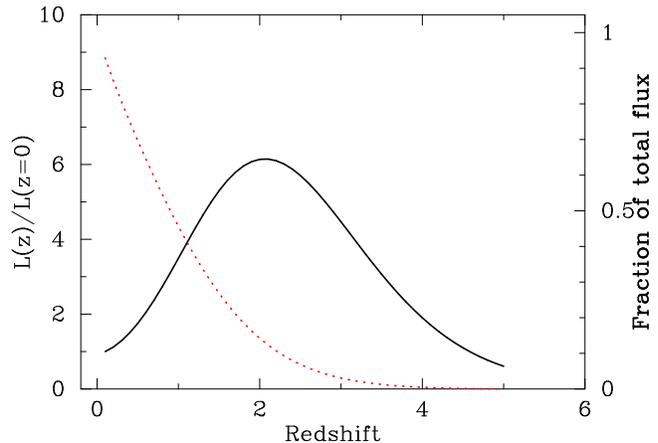}}}
\caption[]{The redshift evolution of the comoving volume emissivity of
AGN in X-rays, $\mathcal{E}(z)$ (used in the calculation of the
scattered fraction of X-rays in Eq.~\ref{eq:fe}) is depicted by the
{\it solid line}. The comoving volume emissivity peaks at $z\sim
2$. The {\it red-dotted line} shows the fraction of the total AGN
contribution to the SXB that was emitted at redshifts $>z$. We find
that $\sim 50\%$ ($10\%$) of the total AGN contribution to the SXB
originates from AGN at $z>1$ ($z>2$). This implies that our method is
most sensitive to dust at redshifts $z\lsim 2$, which is precisely the
range probed by supernova surveys.}
\label{fig:eps}
\end{figure}

Since AGN dominate the SXB, we express the X-ray emissivity in terms
of an integral over the AGN luminosity function,
\begin{equation}
\mathcal{L}(z)=\int_{L_{\rm min}}^{L_{\rm max}}L \psi(L,z) d\log L,
\end{equation} 
where $\psi(L,z)d\log L$ denotes the comoving number density of AGN
with a soft X-ray luminosity (integrated over the photon energy band
between $(0.5$--$2.0)\times (1+z)$ keV) within the interval $\log L
\pm d\log L/2$. For the X-ray luminosity function, $\psi(L,z)d\log L$,
we use the fitting formula of \citet{Hopkins} with $\log L_{\rm
min}=40.4$ and $\log L_{\rm max}$=48.0. The redshift evolution of the
comoving X-ray emissivity, $\mathcal{L}(z)$, is shown in
Fig.~\ref{fig:eps}, where we have normalized $\mathcal{L}(z)$ to the
present-day X-ray emissivity at $z=0$. Figure~\ref{fig:eps} shows that
the comoving emissivity peaks around a redshift $z\sim 2$. Also shown
is the fraction of the total AGN contribution to the soft X-ray
background that was emitted at redshifts $>z$ ({\it red dotted line}):
for example, we find that $\sim 50\%$ ($10\%$) of the total AGN
contribution to the soft X-ray background comes from AGN at $z>1$
($z>2$). This implies that our method is most sensitive to dust at
lower redshift (i.e. $z<2$).

The total soft X-ray background in the 0.5--2.0 keV band that one gets
by integrating the X-ray emissivity of AGN over redshift is
$5.89\times 10^{-12}$ erg s$^{-1}$ cm$^{-2}$ deg$^{-2}$, which
accounts for $\sim 78\%$ of the total background in the 0.5--2.0 keV
band, as calibrated by \citet{M03}.

\section{Constraints from the Unresolved SXB}
\label{sec:result}

In \S~\ref{sec:sxb} we argued that only a small fraction, $F_{\rm
halo} \lsim 10\%$, of the total SXB can be accounted for by X-ray
halos around AGN due to scattering by intergalactic dust. AGN account
for $\sim 80\%$ of the total SXB, and so we require that $F_{\rm
tot}\equiv \int_{0.5\hs{\rm keV}}^{2.0\hs{\rm keV}} F(E) S(E)
dE/\int_{0.5\hs{\rm keV}}^{2.0\hs{\rm keV}} S(E)dE < F_{\rm
halo}\times(100\%/80\%)=12.5\%$, where $F(E)$ is given by
Eq.~(\ref{eq:fe}) and where $S(E)$ denotes the spectral energy density
of the SXB. For simplicity, we take $S(E)\propto E^{-1}$, but point
out that our final results depend only very weakly on this choice. The
constraint $F_{\rm halo}< 10\%$ translates to a constraint on either
$n(z)$ (through Eq.~\ref{eq:tau1}, for a fixed grain size), or on
$\frac{dn}{da}(z)$ (through Eq.~\ref{eq:tau2} for a broad grain size
distribution). Once these quantities have been constrained, we apply
Eq.~(\ref{eq:tau1}) and Eq.~(\ref{eq:tau2}) to constrain the optical
depth of the IGM to optical/infrared photons of observed energy
$E=1.5$ eV ($\lambda=8269$\AA), which characterise supernova
surveys. In this last calculation, we do not use just the scattering
efficiency factor $Q_{\rm scat}$ but rather the {\it total} efficiency
for both scattering and absorption, $Q_{\rm tot}\equiv Q_{\rm
scat}+Q_{\rm abs}$. At the wavelengths of interest, the total
efficiency for optical/infrared photons is $Q_{\rm tot}\approx 2$
regardless of the grain composition \citep[provided that $a\gsim
0.1\mu$m; see Fig.~2-4 of][]{LD93}.
 
We focus on four different models, in which we assume either a single
grain radius of $a=1\mu$m, or a continuous size distribution of the
form $\frac{dn}{da}\propto a^{-3.5}$ with $a_{\rm min}=0.2\mu$m and
$a_{\rm max}=2\mu$m, as implied by interstellar extinction within the
Milky-Way galaxy\citep{MRN}. Our constraints are most effective for
$a\gsim 0.2\mu$m, since the scattering efficiency drops rapidly for
smaller grains at X-ray energies $\gsim 0.5$ keV (for which $Q_{\rm
scat}\propto a^{-2}$ in Eq.~\ref{eq:Q}). On the other hand, grains
larger than $a\sim 2\mu$m scatter X-rays into compact halos with
angular sizes much smaller than $2\sigma\lsim 1(a/2\mu{\rm m})(E/{\rm
keV})^{-1}$ arcmin. Especially photons with an observed energy of 2
keV that were scattered by these grains at higher redshift (say
$z\gsim 2$) would be scattered into halos of radius $< 8$
arcsec. Hence, a significant fraction of the flux in these scattered
X-ray halos may have already been included in the measurement of the
AGN flux.  In addition, we consider either a constant comoving dust
density, or a comoving dust density that decreases with increasing
redshift. In the evolving case, we assume that the comoving dust
density traces the stellar mass density in the Universe, since dust is
a by-product of star formation. In this case, the comoving dust
density is proportional to the integrated star formation rate density,
$n(z) \propto\int _{z}^{\infty}dz'
\frac{\dot{\rho_{*}}(z')}{(1+z')\mathcal{E}(z')}$, where we use the
fitting formula for $\dot{\rho_{*}}$ that was derived by \citet[][
their Eq.~51]{HS03}. Below we discuss our four models individually:

\begin{itemize}

\item {\bf Model I}: {\it $a=1\mu$m and $n$=constant.}
Figure~\ref{fig:dtdz1} shows our upper limit on the opacity of the IGM
to optical/infrared photons of observed energy $E=1.5$ eV
($\lambda=8269$\AA) out to a redshift $z$, $\tau_{\rm GD}(z)$.  The
{\it solid line} shows our upper limit for $F_{\rm halo}=10\%$, while
the {\it blue dashed lines} ({\it red dotted line}) corresponds to
$F_{\rm halo}= 5\%$ ($F_{\rm halo}=15\%$). Figure~\ref{fig:dtdz1}
implies that $\tau_{\rm GD}(z=0.5)\lsim 0.04(F_{\rm halo}/10\%)$ and
$\tau_{\rm GD}(z=2.0)\lsim 0.2(F_{\rm halo}/10\%)$. Our upper limit on
the IGM opacity can be also expressed as an upper limit on the density
parameter in dust (in units of the critical mass density $\rho_{\rm
crit}=1.88 \times 10^{-29}h_{100}^2$ g cm$^{-3}$) of $\Omega_{\rm
d}\equiv [n(\frac{4\pi}{3}\pi a^3) \rho_{gr}]/\rho_{\rm crit}$$\lsim
10^{-4}(F_{\rm halo}/10\%)(\rho_{gr}/3\hs {\rm g}\hs {\rm cm}^{-3})$,
where $\rho_{gr}$ denotes the material density within the dust
grains\footnote{This upper limit on $\Omega_{\rm d}$ is significantly
higher than the upper limit quoted by \citet{P06}. However our
constraints on $\tau_{\rm GD}$ are comparable (see
\S~\ref{sec:discuss}). This discrepancy arises because \citet{P06}
used $Q_{\rm scat}\propto a^2E^{-2}$ for all values of $Q_{\rm scat}$,
which significantly boosts the grain opacity per unit mass.}
\citep[e.g.][and references therein]{Ormel08}.

\noindent
\item {\bf Model II}: {\it $dn/da \propto a^{-3.5}$ and $n$=constant.}
The {\it red dotted line} in Figure~\ref{fig:dtdz2} shows the upper
limit on $\tau_{\rm GD}(z)$ obtained by requiring that $F_{\rm halo}=
10\%$. Our constraints on this model are weaker because the total
opacity is now dominated by the smaller grains which are less
efficient in scattering X-rays. Therefore, the overall number density
of grains needs to be increased in order to produce the same opacity
to X-ray photons; this in turn boosts the allowed opacity to
optical/infrared photons. In this case, we get $\Omega_{\rm d} \lsim
7\times 10^{-5}(F_{\rm halo}/10\%)(\rho_{gr}/3\hs {\rm g}\hs {\rm
cm}^{-3})$.

Our constraint on $\Omega_{\rm d}$ is slightly tighter in this model
for the following reason.  Consider the simplest case in which $Q_{\rm
scat}=1.5$ for all grain sizes of interest. The total mass that is
required to produce a given $\tau_{\rm GD}$ would then scale as
$\propto\rho/\tau_{\rm GD}\propto \int da\frac{dn}{da}a^3/\int
da\frac{dn}{da}a^2 \propto a_{\rm max}^{1/2}a_{\rm min}^{1/2}$ (where
we assumed that $a_{\rm min}\ll a_{\rm max}$). Therefore, decreasing
$a_{\rm min}$ {\it decreases} the mass in dust that is required to
produce a given $\tau_{\rm GD}$. In reality, $Q_{\rm scat}\propto
a^{2}$ for grain size smaller than some threshold value that depends
on the energy of the X-ray photons (Eq.\ref{eq:Q}), which makes the
$a_{\rm min}$-dependence of our constraint on $\Omega_{\rm d}$ a bit
more complicated.

\begin{figure}
\vbox{\centerline{\epsfig{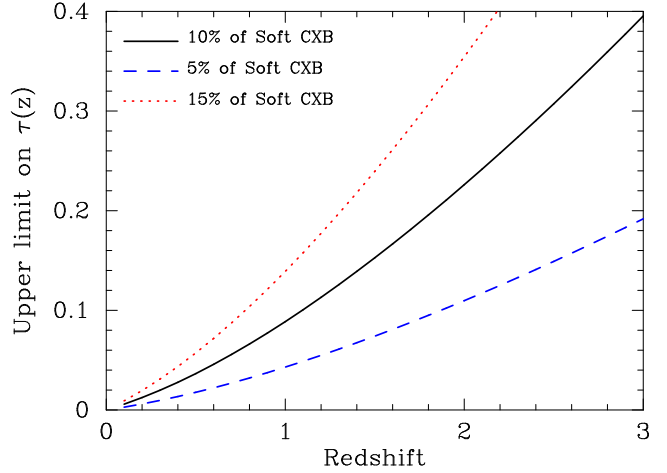}}}
\caption[]{Upper limits on the opacity of the IGM to optical/infrared
radiation (in the observer's frame) out to redshift $z$, $\tau_{\rm
GD}(z)$, due to intergalactic dust grains with a radius of
$a=1\mu$m. The {\it solid line} shows the constraint that is obtained
when we assume that a fraction $F_{\rm halo}=10\%$ of the total SXB is
accounted for by X-ray halos around AGN due to intergalactic dust
scattering. {\it Blue dashed} ({\it red dotted}) {\it lines} show the
upper limits for $F_{\rm halo}=5\%$ ($F_{\rm halo}=15\%$).}
\label{fig:dtdz1}
\end{figure} 

\begin{figure}
\vbox{\centerline{\epsfig{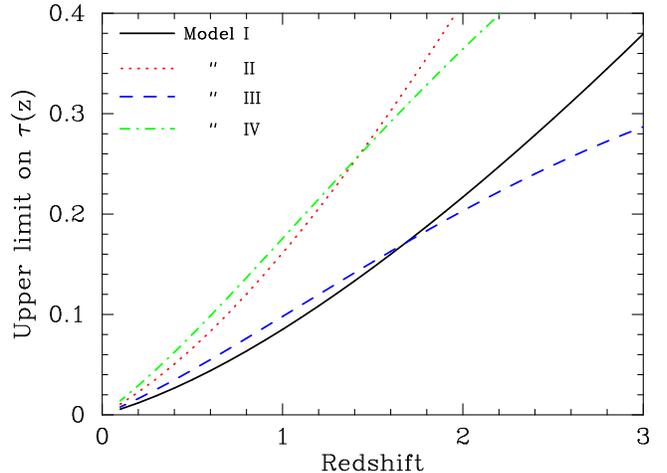}}}
\caption[]{Same as Fig.~\ref{fig:dtdz1} for $F_{\rm halo}=10\%$  and
different choices of the grain size distribution or the redshift
evolution of the intergalactic dust (see \S~\ref{sec:result} for a
detailed description of each model). Our constraint on the IGM opacity
to optical/infrared photons appears to be robust, and exhibits only a
mild dependence on the details of the underlying model.}
\label{fig:dtdz2}
\end{figure} 

\noindent
\item {\bf Model III}: {\it $a=1\mu$m and $n(z)\propto \int
_{z}^{\infty}dz'
\frac{\dot{\rho_{*}}(z')dz'}{(1+z')\mathcal{E}(z')}$.} In this model,
the number density of grains decreases with increasing redshift.  For
example, the number density of grains at $z=1$ ($z=2$, $z=3$) is $0.7$
($0.5$, $0.3$) times its value at $z=0$. For this reason, the IGM
becomes increasingly transparent with increasing redshift compared to
{\bf Model I}, and $\tau_{\rm GD}(z)$ increases more moderately
towards higher redshifts than in {\bf Model I}. For this model, we
find that $\Omega_{\rm d}(z=0)<1.5\times 10^{-4}(F_{\rm
halo}/10\%)(\rho_{gr}/3\hs {\rm g}\hs {\rm cm}^{-3})$. The redshift
dependence of our upper limit on $\Omega_{\rm d}$ scales as $n(z)$.

\noindent
\item {\bf Model IV}: {\it $dn/da \propto a^{-3.5}$ and $n(z)\propto
\int _{z}^{\infty}dz'
\frac{\dot{\rho_{*}}(z')dz'}{(1+z')\mathcal{E}(z')}$.} The differences
between this model and {\bf Model III} are similar to the differences
between {\bf Model I} and {\bf Model II}. We find that $\Omega_{\rm
d}(z=0)\lsim 9\times 10^{-5}(F_{\rm halo}/10\%)(\rho_{gr}/3\hs {\rm
g}\hs {\rm cm}^{-3})$.

\end{itemize}

\section{Discussion}
\label{sec:discuss}
\begin{figure}
\vbox{\centerline{\epsfig{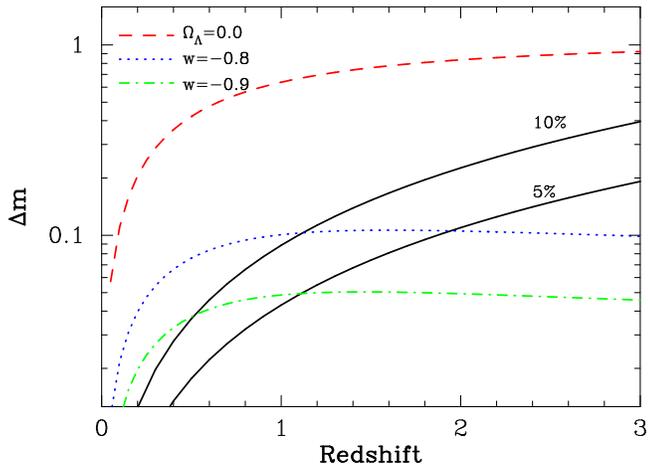}}}
\caption[]{Upper limit on the allowed increase in the apparent
magnitude of optical/infrared sources, $\Delta m(z)$, due to
intergalactic grey dust for $F_{\rm halo}=10\%$ and $5\%$ (black {\it
solid lines}). We also consider the effective dimming of sources as a
result of a departure in the value of the luminosity distance from
that predicted by the standard $\Lambda$CDM cosmology, and compare the
allowed deviation from dust to that caused by different choices of the
equation of state of the dark energy, $w=-0.8$ and -0.9 as well as an
$\Omega_m=1, \Omega_\Lambda=0$ cosmology.}
\label{fig:cosmo1}
\end{figure} 

We next translate our upper limits on the opacity of intergalactic
dust grains to an upper limit on the {\it increase} in the apparent R
and I-band magnitude of optical/IR sources, $\Delta
m(z)=1.086\tau_{\rm GD}(z)$. The black {\it solid lines} in
Figure~\ref{fig:cosmo1} show our upper limit on $\Delta m(z)$ for
$F_{\rm halo}=10\%$ and $5\%$, in comparison to the effective dimming
of sources as a result of a departure in the value of the luminosity
distance from that predicted in the standard $\Lambda$CDM
cosmology. We compare our upper limit to the difference in the
distance modulus (defined as the difference between the apparent and
absolute magnitude of a source) for different cosmological
models\footnote{Our constraint on $\tau_{\rm GD}$ actually depends
weakly on the assumed cosmology. Suppose that the equation of state
of dark energy is measured to be $w\pm \delta w_{\rm stat}+\delta
w_{\rm sys}$, where $\delta w_{\rm stat}$ and $\delta w_{\rm sys}$
denote the statistical and systematic errors, respectively. The
systematic error is related to $\tau_{\rm GD}$ as $\delta w_{\rm
sys}\approx -2(\tau_{\rm GD}+\frac{d\tau_{\rm GD}}{dw}dw)$. We have
verified that $\frac{d\tau_{\rm GD}}{dw}\ll \tau_{\rm GD}$, and the
cosmology dependence of the upper limit on $\tau_{\rm GD}$ is not
important.}.

The {\it red dashed line} shows the extra dimming in the standard
$\Lambda$CDM cosmology relative to the Einstein-de-Sitter Universe
($\Omega_m=1$, $\Omega_{\Lambda}=0$). Clearly, our model upper limits
are well below this line. This implies that our upper limit on the
intergalactic grey dust abundance rules out the 'Einstein-De-Sitter
Universe $+$ grey dust' model for the luminosity distance inferred
from supernova at $z<1$. The {\it blue dotted line} ({\it green
dot-dashed line}) shows the extra dimming that one expects for a
Universe with $\Omega_{\rm m}=0.27$, $\Omega_{\Lambda}=0.73$ and
$w=-0.8$ or -0.9 relative to the standard $\Lambda$CDM. At $z<1$ our
upper limit on the intergalactic opacity with $F_{\rm halo}=10\%$ is
comparable to the difference in the distance modulus for models with
$w\sim -0.8$ and $w=-1$.

Figure~\ref{fig:cosmo1} suggests that at $z<1$ or $z\lsim 0.5$ the
systematic uncertainty that is introduced by grey dust with $F_{\rm
halo}=10\%$ is smaller than the boost in the distance modulus for a
model with $w=-0.8$ or $w=-0.9$, respectively. At higher redshifts,
the possibility that dust mimics the behavior of dark energy with
$w=-0.8$ or $w=-0.9$ for a Universe in which $w=-1.0$, is not ruled
out yet. Also shown for completeness is our upper limit on the amount
of attenuation by grey dust if $F_{\rm halo}<5\%$, which implies that
$\tau_{\rm GD} \lsim 0.05$ at $z<1$.

In \S~\ref{sec:result} we inferred that $\Omega_{d}\lsim
10^{-4}(\rho_{gr}/3\hs {\rm g}\hs {\rm cm}^{-3})$ for all models,
which is comparable to the intergalactic dust density that is
predicted for certain models \citep{A99B,H00}. Suppose that we can
constrain the intergalactic dust abundance down to a level of
$\Omega_{\rm d}\lsim 10^{-5}$, then this would allow us to put
valuable constraints on models of the enrichment of the IGM.
Furthermore, constraints at this level would greatly reduce the
systematic uncertainties introduced by grey dust to a level of
$\tau_{\rm GD}(z=2) \lsim 0.02$. One way to get better constraints on
the intergalactic grey dust abundance from X-ray observations is to
put upper limits on the total X-ray flux in the halos surrounding
individual X-ray point sources \citep[as in][]{P06}. \citet{P06}
obtained an upper limit on $\tau_{\rm GD}(z=4.3)\lsim 0.18$ (assuming
$n$=constant, and $a=1\mu$m, which corresponds to our {\bf Model I}),
which is a factor of $\sim 3$ tighter than our constraint. However, in
difference from our constraints, this result depends strongly on the
assumed grain size \citep[e.g.][]{ML91}: e.g., grains with $a=0.5\mu$m
($a=0.25\mu$) would scatter the X-rays over an area that is four
(sixteen) times larger, which would weaken the upper limit by a
corresponding factor of $\sim 4$ ($\sim 16$). This technique is
nevertheless powerful and can be applied to individual AGN at lower
redshifts. For example, the observed flux from an equally luminous
X-ray source at $z=1$ is $\sim 30$ times larger. Therefore, with an
equally long X-ray observation one would be able to constrain the
surface brightness of the scattering halo to a level that is $\sim 30$
times lower, thus greatly reducing the upper limit on $\tau_{\rm
GD}(z=1)$. Stacking of the X-ray images of luminous nearby sources may
reduce this upper limit even further.

We note that scattering of X-rays by smaller dust grains in the halos
hosting the X-ray sources can also produce halos around sources
\citep{P91,P95}. Furthermore, X-rays may also be scattered by large
dust grains in our own galaxy. However, as long as X-ray halos are not
observed, these additional opacities only render our upper limits more
conservative. The absence of X-ray halos around individual AGN may
therefore also constrain the opacity of the dust in spatially extended
(several tens of kpc) winds that may surround quasars. On the other
hand, if X-ray halos {\it are} detected there might be degeneracies in
their interpretation.

Our constraint that $\tau_{\rm GD}(z=1)\lsim 0.15$ is among the
tightest in the literature. \citet{AH00} showed that the unresolved
fraction of the {\it Far Infrared} Background (at $\lambda=850\mu$m)
can be used to constrain the intergalactic grey dust opacity to a
level $\tau_{\rm GD}(z=0.5)\lsim 0.15$. Their upper limit applies to
dust grains with radii $a_{\rm min}\gsim 0.1\mu$m. \citet{Mr03} were
able to put an upper limit on the total dimming by grey dust at the
level $\Delta m(z=1)\sim \tau_{\rm GD}(z=1)\lsim 0.2$ (99\% confidence
level) by investigating the color evolution in the spectra of $2740$
quasars at $0.5 \lsim z_{\rm qso} \lsim 2.0$.  Furthermore,
\citet{More08} constrained the intergalactic dust opacity to
$\tau_{\rm D}(z=0.35)-\tau_{\rm D}(z=0.20)\lsim 0.13$
(95\% confidence level. Note that the subscript 'D' emphasises that
this constrint refers to dust in general rather than grey dust) by
comparing the luminosity distance to the angular diameter distance
inferred from baryonic acoustic oscillations. This upper limit is
significantly weaker than the upper limit derived in this
paper. However, \citet{More08} anticipate their upper limit to improve
by a factor of $\sim 10$ within the next few years. The advantage of
their test is that it works regardless of the composition of
intergalactic dust (but this may be a disadvantage if one would like to constrain the physical properties of the intergalactic dust).

\section{Conclusions}
\label{sec:conc}

Scattering by dust grains in the Intergalactic Medium (IGM) produces
diffuse X-ray halos around AGN, with a surface brightness that is
typically too faint to be detected. Taking account of the X-ray
emission by star forming galaxies and galaxy clusters, leaves only a
fraction $F_{\rm halo}\lsim 10\%$ of the Soft X-Ray Background (SXB)
to be possibly associated with these diffuse X-ray halos.

The SXB constrains the opacity of the intergalactic 'grey' dust, which
consists of large grains ($a \gsim 0.1\mu$m) that produce little
reddening at optical/infrared wavelengths \citep{A99A}. Thus, grey
dust is a source of systematic uncertainty for supernova surveys that
aim to improve the precision on measuring the redshift dependence of
the luminosity distance, in an attempt to constrain the cosmic
evolution of the equation of state of the dark energy.

Our analysis placed an upper limit on the dust opacity of the IGM to
optical/infrared photons (with energy $E\gsim 1$ eV) of $\tau_{\rm
GD}\lsim 0.15(F_{\rm halo}/10\%)$ to $z\sim 1$ (and $\tau_{\rm
GD}\lsim 0.4[F_{\rm halo}/10\%]$ to $z\sim 2$; see
Fig.~\ref{fig:dtdz1}). Our constraints are most effective for large
dust grains with radii in the range $a=0.2$--$2.0\mu$m. Our quoted
upper limits are only weakly sensitive to the assumed size
distribution of the dust grains within this size range, or to the
precise redshift evolution of the overall dust content (see
Fig.~\ref{fig:dtdz2}).

Significantly improved constraints may be obtained by stacking X-ray
point sources in the redshift interval $z=0$--$2$. This approach has
the potential to eliminate one systematic source of uncertainty for
future supernova surveys which aim to determine the redshift
dependence of the dark energy equation of state at the percent level
of precision. Combining the constraints from X-ray observations with
constraints from the unresolved Far Infrared Background
\citep[][]{AH00}, the reddening\footnote{After the submission of our
paper, \citet{M09} reported a detection of the reddening of quasars by
intergalactic dust out to a few Mpc around galaxies at $z=0.3$. The
observed reddening implies a slope of the extinction curve,
$R_V=3.9\pm 2.6$, which is consistent with that of interstellar dust
(for which $R_V=3.1$). For $R_V=3.1$ the observed reddening
corresponds to a cosmological dust opacity of $\tau_{\rm
D}(z=0.5)=0.03$ and $\tau_{\rm D}(z=1.0)=0.05-0.09$. The reported
uncertainty on $R_V$ allows for a contribution of a grey dust
component to the intergalactic dust opacity of comparable magnitude.}
\citep[][]{Mr03,M09} and comparing the luminosity and angular diameter
distances \citep[][]{More08}, may constrain further the properties of
intergalactic dust. In addition, these techniques might also be used
to search for spatial fluctuations in the intergalactic dust abundance.

\noindent
{\bf Acknowledgments} This research was supported in part by Harvard
University funds. We thank Zolt\'{a}n Haiman, Matthew McQuinn, Adam
Lidz, and Ryan Hickox for helpful discussions.

\label{lastpage}

\end{document}